\documentclass[journal,10pt]{IEEEtran}
\usepackage{subfigure}
\PassOptionsToPackage{subfigure}{tocloft}
\usepackage{CJK}
\usepackage{algorithm}
\usepackage{algpseudocode}
\usepackage{amsmath}
\usepackage{amssymb}
\usepackage{psfrag}
\usepackage{graphicx}
\usepackage{epstopdf}
\usepackage{multirow}
\usepackage{stfloats}
\usepackage{cite}
\usepackage{color}
\usepackage[normalem]{ulem}
\usepackage{arydshln}
\usepackage{enumerate}
\usepackage{bbm}
\usepackage{bm}
\def\BibTeX{{\rm B\kern-.05em{\sc i\kern-.025em b}\kern-.08em T\kern-.1667em\lower.7ex\hbox{E}\kern-.125emX}}
\title{Large Intelligent Surface/Antennas (LISA): Making Reflective Radios Smart}
\author{
Ying-Chang Liang, \emph{Fellow, IEEE}, Ruizhe Long, Qianqian Zhang, Jie Chen, Hei Victor Cheng, and Huayan Guo\\
\thanks{Y.-C. Liang is with the Center for Intelligent Networking and Communications (CINC), University of Electronic Science and Technology of China (UESTC), Chengdu 611731, China (e-mail: liangyc@ieee.org).}
\thanks{R. Long, Q. Zhang, J. Chen and H. Guo are with the National Key Laboratory of Science and Technology on Communications, University of Electronic Science and Technology of China, Chengdu 611731, China, and also with the Center for Intelligent Networking and Communications (CINC), University of Electronic Science and Technology of China, Chengdu 611731, China.}% (e-mail: ruizhelong@gmail.com; qqzhang\_kite@163.com; chenjie.ay@gmail.com; guohuayan@pku.edu.cn)
\thanks{H. V. Cheng is with The Edward S. Rogers Sr. Department of Electrical and Computer Engineering, University of Toronto, Toronto ON M5S 3G4, Canada.}
}

\begin{document}

\maketitle
%\newpage

\begin{abstract}
Large intelligent surface/antennas (LISA), a two-dimensional artificial structure with a large number of reflective-surface/antenna elements, is a promising reflective radio technology to construct programmable wireless environments in a smart way. Specifically, each element of the LISA  adjusts the reflection of the incident electromagnetic waves with unnatural properties, such as negative refraction, perfect absorption, and anomalous reflection, thus the wireless environments can be software-defined according to various design objectives. In this paper, we introduce the reflective radio basics, including backscattering principles, backscatter communication, and reflective relay, and  the fundamentals and implementations of LISA technology. Then, we present an overview of the state-of-the-art research on emerging applications of LISA-aided wireless networks. Finally, the limitations, challenges, and open issues associated with LISA for future wireless applications are discussed.
\end{abstract}


\section{Introduction}
Spectral and energy efficiencies have always been the key considerations in the design of modern wireless communications.
Conventionally, communication systems are generally designed based on the active-radio mechanism, with which the wireless  transmitter generates the transmitted \emph{radio frequency} (RF) signal using a series of active components such as \emph{digital-to-analog converter} (DAC), local oscillator, up-converter, and \emph{power amplifier} (PA).
With more and more {\emph{base stations}} (BSs) and wireless terminals being deployed to support the exponentially growing mobile traffic and massive wireless connections, the power and spectrum consumption as well as implementation cost will increase to the prohibitive levels.
Recently, reflective radio technology has emerged as an attractive solution for designing next-generation spectral and energy efficient communication systems. Specifically,  the \emph{reflective devices} (RDs), which do not use expensive and  power-hungry active components, could transmit signals to their receivers or enhance the transmission of other primary communication systems  using {\emph{electromagnetic}} (EM) scattering principles. The former is called \emph{backscatter communication} (BSC), and the later is referred to as \emph{reflective relay}. %Fig.\ref{roadmap} shows the technology evolving path of various reflective radio techniques.

\begin{figure}
[t]
\centering
\includegraphics[width=0.8\columnwidth]{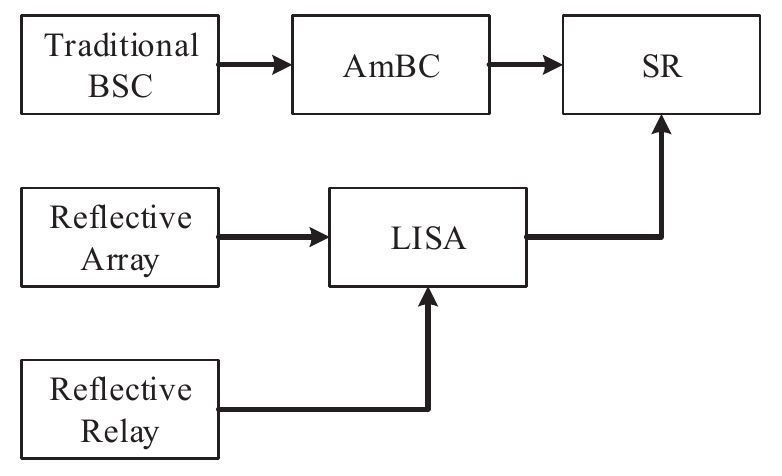}
\caption{Technology evolving path of the reflective radio techniques.}
\label{roadmap}
%\vspace{-1.5em}
\end{figure}

BSC has been widely used to support low-power communications for {\emph{Internet of Things}} (IoT) \cite{Boyer2014Backscatter,griffin2009complete,Kimionis2014bistatic}. Traditional BSC, either monostatic or bistatic, relies on a continuous carrier wave signal generated by a dedicated RF emitter separated from the RD, which transmits its messages to the receiver using backscatter modulation. \emph{Ambient backscatter communication} (AmBC) exploits RF signals from the primary source such as TV tower, cellular BS, and WiFi access point in the ambient environment, \cite{liu2013ambient,parks2015turbocharging,VanHuynh2018}, thus no dedicated RF emitter is required. In addition, the AmBC system naturally shares the same spectrum with the primary system, resulting in better spectrum utilization \cite{Niyato2017RFCRN,KangX2017RidingPrimary,xiaos2019BackFi,GaoFF2017Noncoherent,guo2019exploiting,YangG2017OFDM}.

Reflective relay has been used to enhance the {\emph{quality-of-service}} (QoS) of users in a primary system suffering from unfavorable propagation conditions \cite{shahbazi2012improving,shahbazi2011placement}. Conventionally, single antenna is deployed at the reflective relay, leading to a weak reflective link. Reflective array with a large number of antennas has been proposed to enhance the strength of the reflective link \cite{berry1963,patent1987}. More recently, reflective relay has been evolved to \emph{large intelligent surface/antennas} (LISA), or called intelligent reflective surface (IRS), a two-dimensional structure with a large number of reflective-radio elements, each of which could adjust the reflection of the incident electromagnetic waves with unnatural properties, such as negative refraction, perfect absorption, and anomalous reflection.
With  LISA, wireless environments can be software-defined according to various design objectives.
For example, LISA can enable the reflected signals being coherently added at the receiver without introducing additional noise. Thus, the received \emph{signal-to-noise ratio} (SNR) scales up with the square of the number of reflective-radio elements. %Compared with classic {\emph{maximal ratio transmission}} (MRT) and {\emph{maximal ration combining}} (MRC) where the received SNR is scaled up linearly with the number of antenna,
Therefore, LISA achieves a significant SNR gain when the number of reflective-radio elements becomes large.

\emph{Symbiotic radio} (SR) is another novel system generalized from BSC and reflective relay \cite{long2018symbiotic,GuoAccessSR2019,rzlAccess,zhangqqAccess}. Basically, BSC system focuses on the information transmission of RDs, while conventional reflective relay system is designed to assist the transmission of the primary systems. In a SR system, the RD transmits information to its receiver by riding on the RF signal of a primary system, simultaneously, it can enhance the transmission of the primary system. %Therefore, the SR system combines the advantages of BSC and the LISA technology, which greatly improves the spectral and energy efficiencies.

As a summary, Fig.\ref{roadmap}  shows the technology evolving path of various reflective radio technologies.
In this paper, we focus on the LISA technology and provide an
 overview of the relationship between LISA, BSC, AmBC, and SR. We first give a brief introduction on reflective radio, including the backscattering principle, BSC, AmBC, SR, and reflective relay. Then we introduce LISA and analyze its performance gain as well as its implementation. After that, the state-of-the-art research on the applications of LISA in wireless communications will be reviewed. Finally, open issues and future research directions in this emerging field will be pointed out.

\begin{figure}[t]
    \centering\includegraphics[width=0.9\columnwidth]{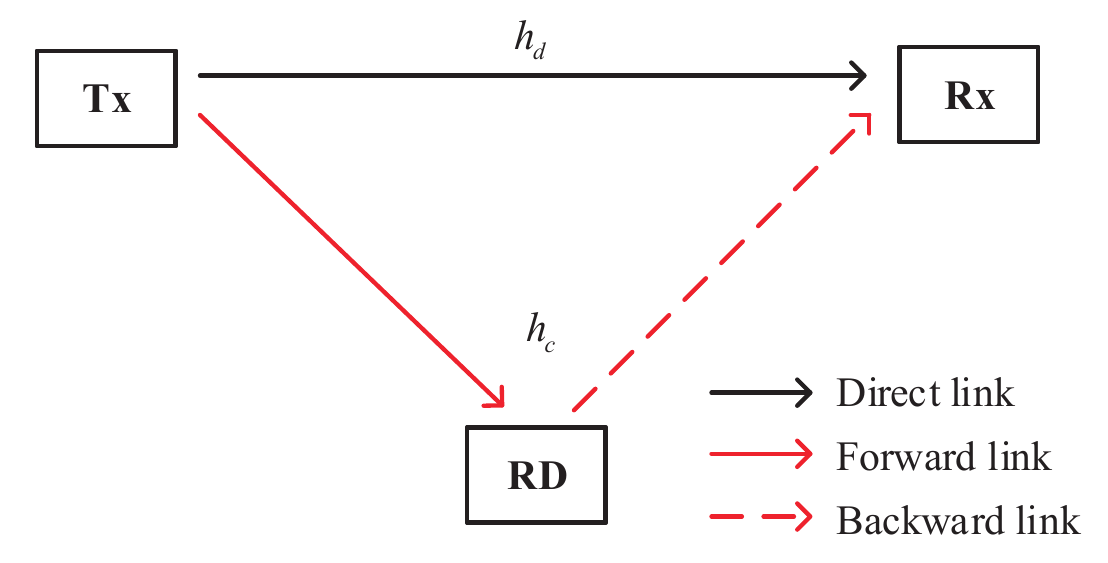}
    \caption{A general reflective radio model consisting of three nodes: a transmitter (Tx), a reflective device (RD), and a receiver (Rx).}\label{fig:PassiveRadio}
%    \vspace{-0.1cm}
\end{figure}

\section{Reflective Radio}

A general model for reflective radio is shown in Fig. \ref{fig:PassiveRadio}, in which the \emph{transmitter} (Tx) transmits signals to the \emph{receiver} (Rx), and the RD transmits its own messages to RX through backscatter modulation or assists the information transmission from the Tx to the Rx. In general, the received baseband signal $y$ at the Rx can be written as:
\begin{equation}\label{eq:GeneralModel}
  y = \underbrace{h_d x}_{\text{direct-link}} + \underbrace{h_c s x}_{\text{backscattered}} +u,
\end{equation}
where $x$ is the transmit signal from the Tx with $\mathbb{E}[x^2]=p_t$ being the transmit power, $s$ is the information symbol at the RD and $u$ is complex Gaussian noise at the Rx with zero mean and variance $\sigma^2_u$; $h_d$ is the direct link channel from the Tx to the Rx and $h_c$ is the reflective channel, composed of a forward channel from the Tx to the RD and a backward channel from the RD to the Rx. Specifically, the received signal can be divided into two signal components: the direct-link signal and the backscattered signal.

Before going deep into the models for each reflective radio technology, we first review the fundamental principles of backscattering which are the basics of reflective radio.
%\begin{figure}[t]
%    \centering\includegraphics[width=1.0\columnwidth]{systemmodel.pdf}
%    \caption{A bistatic backscatter communication system consisting of three nodes: a transmitter (Tx), a reflective device (RD), and a receiver (Rx).}\label{fig:PassiveRadio}
%%    \vspace{-0.1cm}
%\end{figure}

\begin{figure}[t]
    \centering\includegraphics[width=.4\columnwidth]{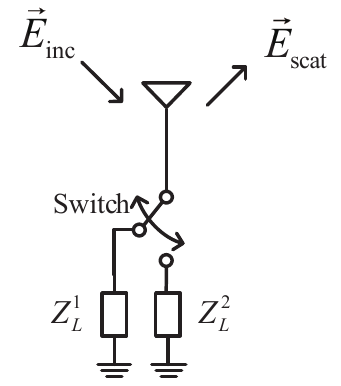}
    \caption{Basic backscattering circuit.}\label{fig:Backscatter}
%    \vspace{-0.1cm}
\end{figure}

\subsection{Principles of backscattering}

To understand how the backscattering works, we exploit Green's decomposition \cite{hansen1989relationships} to decompose the backscattered signal into two components: \emph{structural mode} scattering component and \emph{antenna mode} scattering component. Based on a basic backscattering circuit shown in Fig. \ref{fig:Backscatter}, the scattered field for any load impedance $Z_{L}^{i}$ and a fixed antenna impedance $Z_\mathrm{a}$ can be written as:
\begin{equation}\label{eq:backscatter_mode}
\vec{E}_\mathrm{scat}(Z_{L}^{i})= \vec{E}_\mathrm{scat}(Z^{*}_\mathrm{a}) -\Gamma_i \frac{ I_{o} Z_\mathrm{a}}{2R_\mathrm{a}}\frac{\vec{E}_\mathrm{ant}}{I_\mathrm{a}},
\end{equation}
where the first and second terms on the right hand side of \eqref{eq:backscatter_mode} correspond to the structural and antenna mode components, respectively. The structural mode scattering $\vec{E}_\mathrm{scat}(Z^{*}_\mathrm{a})$ depends on the geometrical layout of the RD antenna and the EM properties of the material. Its value can be measured when the load impedance is matched to the conjugate of the antenna impedance, that is $Z_{L}=Z^{*}_{\mathrm{a}}$. On the contrary, the antenna mode scattering is mainly affected by the load impedance through the reflection coefficient $\Gamma_i$ which is defined as:
\begin{equation}\label{eq:reflection}
\Gamma_i = \frac{Z_{L}^{i}-Z^{*}_{\mathrm{a}}}{Z_{L}^{i} + Z_{\mathrm{a}} }.
\end{equation}
Note the fraction $\frac{ I_{o} Z_\mathrm{a}}{2R_\mathrm{a}}$ in \eqref{eq:backscatter_mode} is the scattering current of the conjugate of the antenna impedance, $\vec{E}_\mathrm{ant}$ is the field radiated by the antenna when the current at the terminal  is $I_\mathrm{a}$ and no externally incident wave is applied \cite{Fuschini2008-p33-35}.

From \eqref{eq:backscatter_mode}, it is observed that the backscattered signal can be adjusted via changing the reflection coefficients in the RD. Thus, the RD can modulate its information or change the phase of the incident signal into a desired one by varying its load impedance.

%\emph{Radar cross section} (RCS) is another effective method to study the signal scattering feature of the RD. A larger RCS indicates that more signal power is reflected. Quantitatively, RCS is calculated in EM analysis as
%\begin{equation}\label{eq:defRCS}
%\sigma=\lim_{r\rightarrow \infty} 4\pi r^2\frac{\left|\vec{E}_\mathrm{scat}\right|^2}{\left|\vec{E}_\mathrm{inc}\right|^2},
%\end{equation}
%where $\vec{E}_\mathrm{scat}$ and $\vec{E}_\mathrm{inc}$ are the far field scattered and incident electric field intensities, respectively \cite{balanis1999advanced}. Considering the BSC applications, the RCS can be expressed by:
%\begin{equation}\label{eq:rfidRCS}
%\sigma_i=\frac{\lambda^2}{4 \pi}G^{2}_{B}\left|A_{s}-\Gamma_{i}\right|^2,
%\end{equation}
%where $G_{B}$ is the RD antenna's gain, $\lambda$ is the wavelength of the radiation and $\Gamma_{i}$ is defined in \eqref{eq:reflection}. Note $A_{s}$ is a complex-value that represents the structural mode component. RCS can help to simplify the formula derivations related to the backscattered signal power in passive radio. Based on RCS, some power constraints can be expressed concisely \cite{Nikitin2007-p431-432,Bletsas2010-p1502-1509}.

\subsection{Traditional Backscatter Communication}
In traditional BSC, the RD exploits backscatter modulation to transmit its messages to the Rx by riding on the \emph{continuous wave} (CW) signal generated by a dedicated RF emitter. By varying the load impedance $Z_{L}$, the RD presents different reflection coefficients, and thus modulates its own information symbol over the CW wave. The Rx extracts the RD's information $s$ from the backscattered signal by recovering the reflection states when the Tx transmit signal $x$ is fixed and known at the Rx (i.e., $x=\sqrt{p_t}$). In this case, the received signal after the direct-link signal cancellation is written as follows,
\begin{equation}\label{eq:BSCmodel}
y = \sqrt{p_t}h_c s + u.
\end{equation}
Take a simple binary modulation as an example, to transmit bits '0' and '1', the RD only needs two states related to the reflection set $\mathcal{A}\in \{\Gamma_1,\Gamma_2\}$. In order to improve the transmission efficiency, high order modulation such as \emph{quadrature amplitude modulation} (QAM) can be applied. Basically, the number of the discrete load impedance states attributes to the RD's modulation schemes. That is, the more reflection states, the higher the modulation order. The key issue of backscatter modulation is to design an appropriate reflection set corresponding to the signal constellation. The relationship between the complex constellation point $s_i \in \mathcal{S}$ and the reflection coefficient $\Gamma_i$ is given by \cite{Thomas2012Quadrature}
\begin{equation}\label{eq:backCon}
\Gamma_i = \alpha \cdot \frac{s_i}{\max_{s\in\mathcal{S}}|s|},
\end{equation}
where $0\leq \alpha \leq 1$ is a constant scalar which is related to power reflection and power transmission coefficients. Once the reflection coefficient $\Gamma_i$ is given, the corresponding load impedance $Z_{L}^{i}$ is determined as
\begin{equation}\label{eq:loadIm}
Z_{L}^{i}=\frac{Z_\mathrm{a}^{*}+\Gamma_i Z_\mathrm{a}}{1-\Gamma_i}.
\end{equation}

\subsection{Ambient Backscatter Communication}

The deployment cost and radio spectrum requirement for traditional BSC may become unaffordable with the exponential growth of the IoT devices.
To overcome this challenging issue, AmBC has been proposed recently, which exploits the RF signals from a primary communication system in the ambient environment \cite{liu2013ambient,parks2015turbocharging,VanHuynh2018}. With AmBC, no dedicated RF emitter is required, which enables more flexible and efficient network deployment. In addition, the backscatter system naturally shares the same spectrum with the primary system, resulting in better spectrum utilization \cite{Niyato2017RFCRN,KangX2017RidingPrimary,xiaos2019BackFi}. In this case, the received signal is written as:
\begin{equation}\label{eq:AmBCmodel}
  y = h_d x + h_csx +u,
\end{equation}
and the RX generally has to decode the RD's symbol $s$ in the presence of unknown ambient signal $x$. Nevertheless, due to the lack of cooperation with the primary system, AmBC suffers from severe {\emph{direct-link interference}} (DLI), and the coverage range and transmission rate of AmBC are limited \cite{GaoFF2017Noncoherent,guo2019exploiting,YangG2017OFDM}.

A promising solution to suppress the direct-link interference in AmBC
   is to deploy a cooperative receiver, which jointly recovers both primary and RD's signals  \cite{YangG2018cooperative,zhang2019constellation,Guo2019WCL}. With such cooperation, a new concept named {\emph{symbiotic radio}} (SR) is proposed \cite{GuoAccessSR2019,rzlAccess,zhangqqAccess}. An interesting phenomenon that has been observed in SR is that when the RD's symbol duration is much longer than that of the primary signal, the backscattered signal may even enhance the transmission of the primary system \cite{long2018symbiotic}. Thus, it becomes mutually beneficial for the IoT transmission to coexist with cellular communications using SR technology.

\subsection{Reflective Relay}

Another important reflective radio technology is the reflective relay, where RD assists the information transmission from the Tx to the Rx. In this case, the direct-link is typically blocked due to obstacles and thus the QoS of the users are suffering from unfavorable propagation conditions \cite{shahbazi2011placement,shahbazi2012improving}. The RD is thus designed to backscatter the incident signal from the Tx as much as possible, providing a reflective transmission path towards the Rx. In this case, the Rx aims to decode the Tx information $x$ when the backscattered symbol $s$ is chosen as a fixed coefficient, thus the received signal can be written as:
\begin{equation}\label{eq:RRModel}
  y = h_c s x + u.
\end{equation}
Let us consider a typical backscatter link budget model for reflective relay \cite{griffin2009complete}. In this model, the received power $P_{R}$ of the backscattered signal is given by
\begin{equation}\label{eq:biLink}
P_{R}= \frac{P_{T}G_{T} G_{R} G^2_{B} X_f X_b M_Bc^4 }{\left(4\pi \right)^4 r^2_f r^2_b\Theta^2 B_f B_b F_{\beta}f^4},
\end{equation}
where $P_T$ is the power transmitted from the Tx, $G_{T}$, $G_R$ and $G_{B}$ are the gains of the Tx antenna, the Rx antenna and the RD antenna, respectively; $X_f$ and $X_b$ are the link polarization mismatch parameters of the forward and backward links, respectively; $M_B$ is the modulation factor at the RD, $c$ is light speed; $r_f$ and $r_b$ denote the distances of the forward and backward links, respectively; $\Theta$ is a gain penalty due to the material attachment; $B_f$ and $B_b$ are path blockage losses of the forward and backward links, respectively; $F_{\beta}$ is the fade marginal; and $f$ is the carrier frequency.

From~\eqref{eq:biLink}, it is observed that the received power $P_{R}$ decreases with the carrier frequency in the power of four, which is different from the general free space propagation model where the received power decreases with the carrier frequency in the power of two. That is, the backscatter link is very weak especially when  the carrier frequency increases. To overcome such drawbacks, a promising solution is to deploy a large number of reflective-radio elements at the RD to enhance the reflective radio performance, which leads to the idea of {\emph{Large intelligent surface/antennas}} (LISA) that we are going to talk about in the next section.

\section{Large Intelligent Surface/Antennas}

{\emph{Large intelligent surface/antennas}} (LISA) is a two-dimensional artificial structure comprising of a large number of passive-radio meta-atoms/antennas with unique properties for the incident EM waves, such as negative reflection/refraction, perfect absorption, and anomalous reflection. Most existing works exploit LISA to steer the incident RF signals to the desired directions resembling a tunable reflective array.
The concept of reflective array was first developed by Berry in 1963 \cite{berry1963}, where open-ended wave-guides were used as the antenna elements. Adjusting the length of the wave-guides corresponds to changing the phase of the re-radiated signal and forming the desired beams.
Interests in reflective arrays emerged in the 1990s, when micro-strip patch antennas were first used for implementing planar arrays \cite{patent1987}. There have been many conceptual proposals for tunable reflective array either based on mechanically-tuned or electrically-tuned antennas. However, demonstrated systems through real hardware have been made available only in recent years, with the tremendous advancements in meta-materials and fabrication technologies.

Compared with traditionally used parabolic reflectors with limited tuning capabilities, the key features of the modern LISA are that the beams can be redirected at any arbitrary angles due to the use of meta-materials that can have a negative reflection index. In addition, modern LISA can be tuned electronically in real time. By exploiting these properties, the implementation of LISA can be integrated into the environments, e.g., walls, ceilings, and buildings. Therefore, LISA has recently been considered as the key enabler for controlling the behavior of wireless environments in next-generation's wireless systems.

The most attractive application of LISA is to serve as reflective relays, where LISA is deployed on the facade of obstacles or cell edge for the purpose of enhancing the QoS of users suffering from unfavorable propagation conditions. In this case, LISA resembles a full-duplex multi-antenna {\emph{amplify-and-forward}} (AF) relay. However, in contrast to the active counterpart, the reflection mechanism of LISA brings several advantages, such as ultra-low power consumption, no self-interference, and no additional thermal noise added to the forwarded signal.

With the use of a large number of reflective-radio meta-atoms/antennas, LISA can do much more than just being a reflective relay. For instance,  LISA could be exploited for interference cancellation, thus assists spectrum sharing in cognitive radio networks and/or increases the secrecy capacity in terms of physical layer security \cite{liaskos2018new,cui2014coding}. Moreover, LISA may collect and focus EM power in the environment by exploiting its large aperture, and thus acts as an auxiliary wireless power source for energy-harvesting devices.

\begin{figure}[t]
    \centering\includegraphics[width=.7\columnwidth]{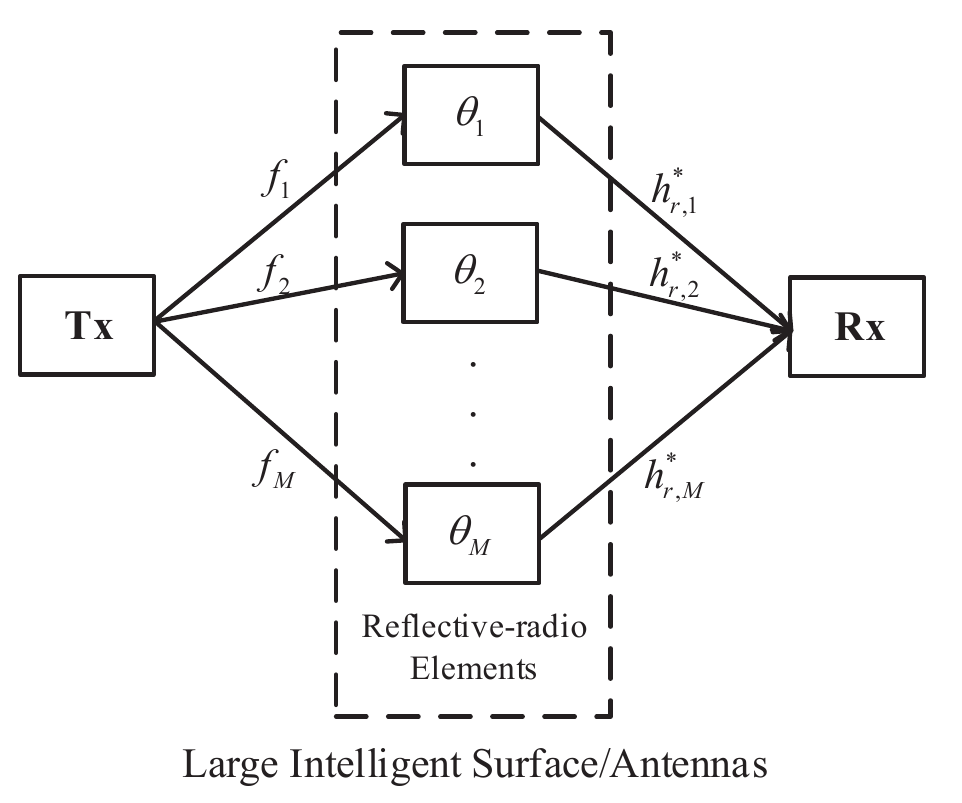}
    \caption{Large intelligent surface/antennas.}\label{fig:LIS}
%    \vspace{-0.1cm}
\end{figure}

\subsection{Performance Gain Analysis for LISA}
In this subsection, we investigate the performance of LISA which deploys a large number of reflective antenna elements. As shown in Fig.~\ref{fig:LIS}, the transmit signal first goes through the LISA with $M$ antennas, then reaches the Rx. The received signal $y$ at the RX is
\begin{equation}\label{eq:LIS_Model}
  y  = \sum_{m=1}^{M} h_{r,m}^{*}\theta_m f_mx + u,
\end{equation}
where $f_m$ is the forward link channel from the Tx to $m$-th reflective-radio element, $h_{r,m}$ is the backward link channel from $m$-th reflective-radio element to the Rx, $\theta_m$ is the reflection coefficient of $m$-th reflective-radio element.
%\subsubsection{Constraints on Reflecting Coefficients}
Generally speaking, there are three different assumptions for the  reflection coefficients, i.e.,
\begin{itemize}
\item \emph{Continuous amplitude and phase-shift:}
\begin{eqnarray}
{\bm {\Phi}}_1 = \left\{ {{\theta }_m\left| {{{\left| {{\theta }}_m \right|}^2} \le 1} \right.} \right\},
\end{eqnarray}
where both the amplitude and phase-shift can be designed for improving the system performance. This is the ideal assumption studied in \cite{guo2019weighted,chen2019IRSPHY}.
\item \emph{Constant amplitude and continuous phase-shift:}
\begin{eqnarray}
{\bm {\Phi}}_2 = \left\{ {\theta_m\left|\theta_m={e^{j\varphi_m}},\varphi_m\in\left[0,2\pi\right)  \right.} \right\}.
\end{eqnarray}
where only the phase-shift can be designed for improving the system performance. This assumption is wildly adopted in existing works  \cite{Ruizhang2018Sufface,huang2018achievable,huang2018large,nadeem2019largearxiv}.
\item \emph{Constant amplitude and discrete phase-shift:} We have
\begin{eqnarray}
\! \!\! \!\! \!\! {\bm {\Phi}}_3\! =\! \!\left\{ {\theta_m\left|\theta_m\!=\!{e^{j\varphi_m}},\varphi_m\!\in\!\!\left\{0,\textstyle{\frac{2\pi}{Q}},\cdots,\textstyle{\frac{2\pi(Q-1)}{Q}}\right\}  \right.} \right\},
\end{eqnarray}
where $Q$ is the number of quantized  reflection coefficient values of the reflective-radio elements on the LISA \cite{yu2019miso,wu2018beamformingOptimization,tan2018enabling,han2018large}.
This assumption is more practical, since in practice, it is costly to achieve continuous  reflection coefficient on the reflective-radio elements due to the hardware limitation\footnote{Note that investigating the system performance with ${\bm \Phi}_1$ and ${\bm \Phi}_2$  is also important since  it serves as the upper bound to that with ${\bm \Phi}_3$.}.
%
%Specifically,  a downlink power minimization problem subject to SNR constraint at a single user by adopting ${\bm {\Phi}}_3$ was investigated in \cite{wu2018beamformingOptimization}, where a low-complexity reflecting coefficient design method was proposed and the performance loss caused by adopting discrete reflecting coefficients was analyzed.

\end{itemize}

%In practice, it is costly to achieve continuous reflecting coefficient on the reflective-radio elements due to the hardware limitation. Hence, it is more practical to apply the discrete reflecting coefficient on the reflective-radio elements, i.e., ${\bm \Phi}_3$, than applying the continuous reflecting coefficients, i.e., ${\bm \Phi}_1$ and ${\bm \Phi}_2$.

All the involved channels $f_m,~h_{r,m},~m=1,2,...,M$ are assumed to follow independent and identically distributed (i.i.d) distribution. Thus, the composite channels $h_{c,m}=f_{m}h_{r,m}^{*},~m=1,2,...,M$ also follow i.i.d distribution. Notice that the backscattered noise is practically negligible as compared to the Rx noise \cite{Bousquet20124}, and thus the noise introduced by backscattering is ignored. With \emph{channel state information} (CSI), the LISA can intelligently adjust its reflection coefficient to deliberately make the backscattered signals coherently added at the Rx, i.e., $\angle\theta_m=-\angle h_{c,m}$. Without loss of generality, we assume that $|\theta_m|=1$ and thus the received signal is written as
\begin{equation}\label{eq:MaximalReflection}
y = \sum_{m=1}^{M}|h_{c,m}|x  + u.
\end{equation}
With the assistance of LISA, the received SNR $\gamma$ now is%\emph{signal-to-noise ratio} (SNR)
\begin{eqnarray}
&\gamma &= \frac{\mathbb{E}\left[(\sum\limits_{m=1}^{M}|h_{c,m}|)^2x^2\right]}{\sigma^2_{u}}\nonumber\\
&&= \frac{p_t\left(\sum\limits_{m=1}^{M}\mathbb{E}[|h_{c,m}|^2]+\sum\limits_{m=1}^{M}\sum\limits_{n\neq m}^{M}\mathbb{E}[|h_{c,m}|]\mathbb{E}[|h_{c,n}|]\right)}{\sigma^2_{u}}\nonumber\\
&&= \frac{p_t\left(M(\sigma^2_c+\mu_c^2)+M(M-1)\mu_c^2\right)}{\sigma^2_{u}} \nonumber\\
&&=\frac{p_t(M^2\mu_c^2+M\sigma^2_c)}{\sigma^2_{u}},
\end{eqnarray}
%where $\mu_c$ and $\sigma_c^2$ are the mean and variance of the random variable $|h_{c,m}|$, respectively. It is observed that the received SNR scales with square of the number of reflective-radio elements, i.e., $M^2$. This is due to the fact that LISA enables the multi-path signals to be added constructively without introducing additional noise. Compared with classic \emph{maximal ratio transmission} (MRT) and \emph{maximal ration combining} (MRC) where the received SNR is scaled up linearly with the number of antenna, i.e., $M$, LISA achieves significant higher SNR improvement when the number of the reflective-radio elements $M$ dramatically increases.
where $\mu_c$ and $\sigma_c^2$ are the mean and variance of the random variable $|h_{c,m}|$, respectively. It is observed that the received SNR scales up with the square of the number of reflective-radio elements, i.e., $M^2$. This is due to the fact that LISA enables the multi-path signals to be added constructively without introducing additional noise. Compared with classic {\emph{maximal ratio transmission}} (MRT) and {\emph{maximal ration combining}} (MRC) where the received SNR is scaled up linearly with the number of antennas, i.e., $M$, LISA achieves a significant SNR gain when the number of the reflective-radio elements becomes large.
\begin{figure}[t]
\centering \includegraphics[width=0.5\textwidth]{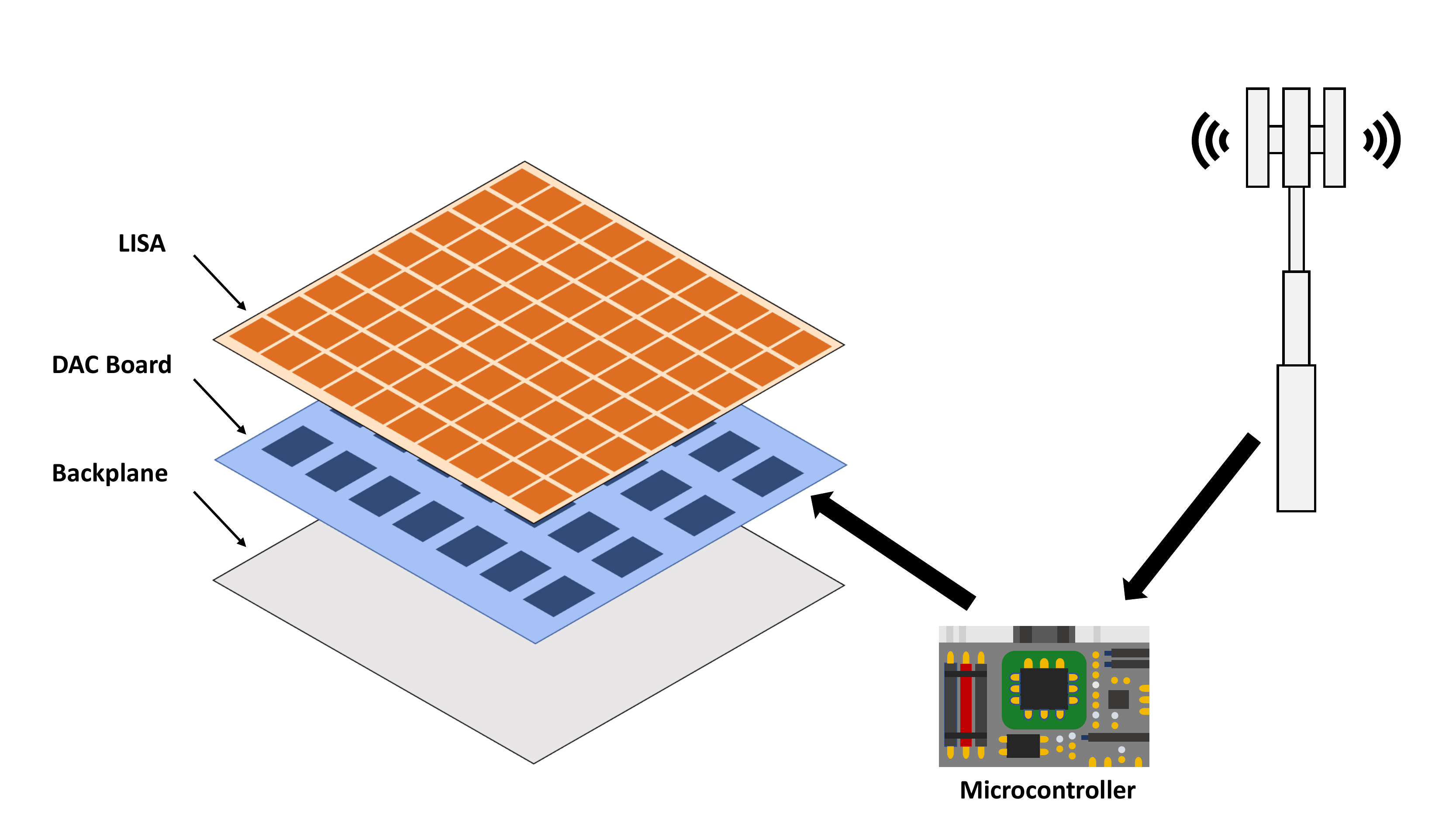}
\caption {A programmable LISA with a micro-controller connected to the access point.} \label{irs_imple}
\end{figure}

\subsection{Implementation of LISA}
%Recently, there has been significant progress in the development and application of reconfigurable antennas, driven by the increasing need for adaptive and multi-functional beam-forming in radar and communication systems. Researchers have become interested in electronically tunable versions of reflect-arrays to realize real-time reconfigurable beam-forming.
The main idea of the LISA is to make the scatterers in the aperture electronically tunable, which can be achieved through introducing discrete elements such as varactor diodes, \emph{positive-intrinsic negative} (PIN) diode switches, ferro-electric devices, and \emph{micro-electromechanical system} (MEMS) switches within the scatterer. As a result, the whole surface can be electronically shaped to adaptively synthesize a large range of beam patterns. At high frequencies, tunable electromagnetic materials such as ferro-electric films, liquid crystals, and new materials such as graphene can be used as the reflective elements to achieve the same effect with lower cost. This has enabled LISA to become powerful beamforming tools in recent years. The main reason is that LISA combines the best features of both reflective antennas and antenna arrays in traditional systems. Thus LISA offers simple, low cost and high-gain as the reflective antennas, meanwhile provides fast, adaptive beam-forming capabilities as the antenna arrays.

\begin{figure}[t]
\centering \includegraphics[width=0.5\textwidth]{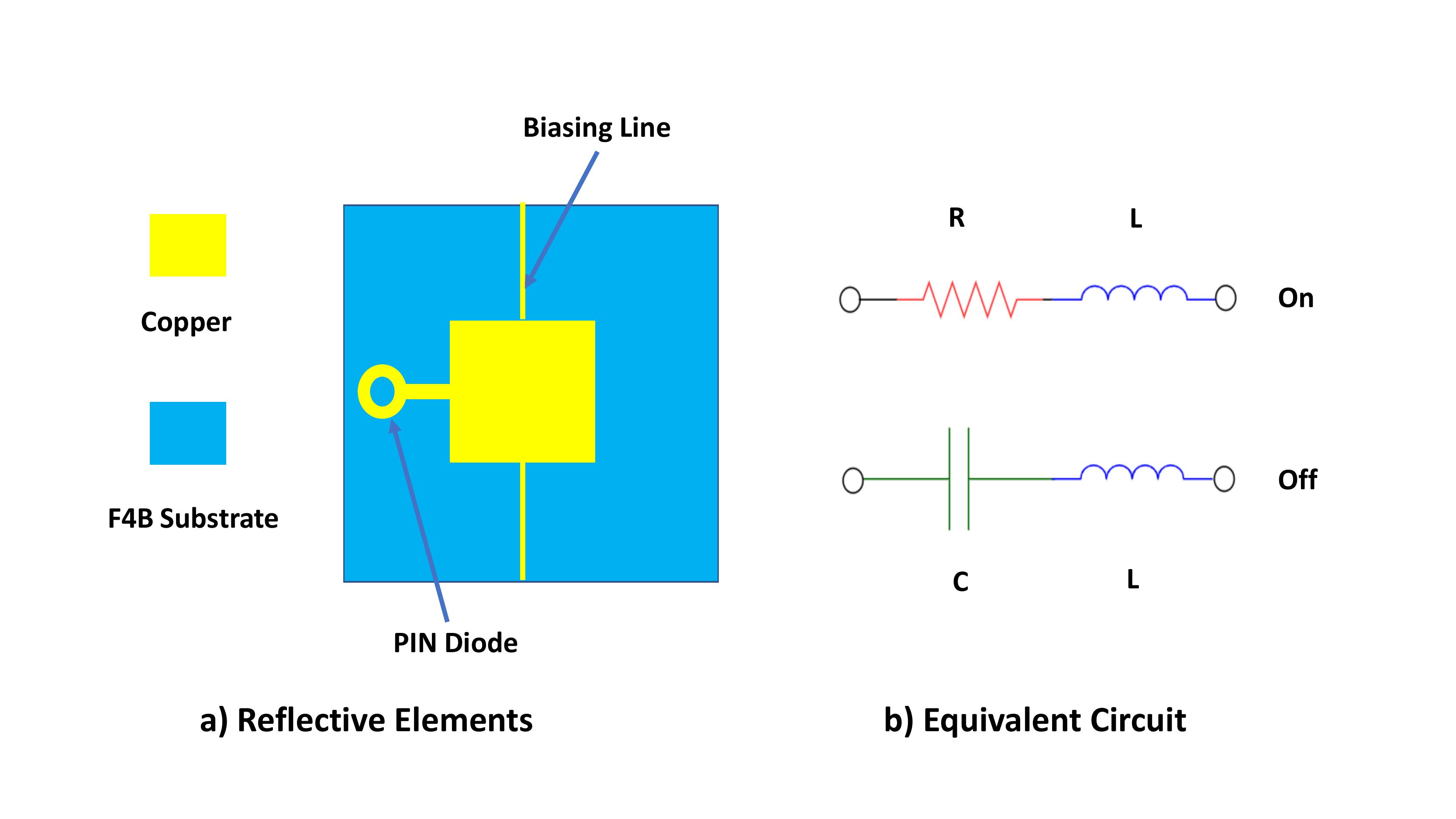}
\caption {An example of the design of reflective element. a) Top view of the reflective element showing the materials and components. b) An equivalent circuit model of the PIN diode at ``On'' and ``Off'' mode. } \label{ref_element}
\end{figure}

Fig. \ref{irs_imple} shows a typical implementation of the LISA systems. It consists of the LISA, the DAC board for controlling the amplitudes and phases of the LISA, and a copper backplane to avoid signal leakage to the surrounding environment. The DAC board works together with the LISA as well as a small micro-controller board for communications between the LISA and the access point, where all the beam control is implemented. The link between the access point and the LISA can be implemented using wired or wireless communication technologies.

\subsection{Reflective Elements}

For the reflective elements in LISA, there exist several different designs. For example, in \cite{tan2018enabling} electronically-controlled relay switches are used to load micro-strip patches. Alternatives include varactor diodes \cite{varactor_diodes2002} and MEMS, where the reflector can be electronically controlled to change the resonance frequency and thus create the desired phase shift and amplitude change \cite{hum_review2014}. Moreover, the rapid advances in meta-surface provides a new design based on liquid crystal \cite{liquid_crystal2017}, where the main idea is to control the DC voltage across the surface to configure the desired direction of the reflected wave in real time. Finally, another notable implementation is through digitally programmable meta-materials \cite{cui2014coding} to manipulate the electro-magnetic waves and thus changing the reflected beams.

Fig. \ref{ref_element} shows an example design of a reflective element and the equivalent circuit model for such design from. When the voltage applied on the PIN diode via the biasing line changes, the PIN diode can switch between the states of ``On'' and ``Off''. To provide a phase difference of 180$^\circ$, the dimensions of the reflective element and the copper patch needs to be carefully designed. In principle, more different phases can be achieved using the same architecture by varying the biasing voltage. Moreover, the amplitude of the reflected waves can also be controlled by adding variable resistors in the reflective elements. Changing the values of the resistor lead to different reflected amplitudes in the range of $[0,1]$. Ideally, we want the amplitudes and phases to be controlled independently and there are already successful designs for such systems \cite{yang2017}. Currently, the technology of meta-surface is undergoing rapid developments and reflective elements with more level of phase/amplitude changes can be expected in the near future.

\section{Applications of LISA}
As aforementioned, LISA has the capability to achieve high spectral and energy efficiency through programming the wireless environment \cite{liaskos2018new,cui2014coding}.
Specifically, LISA can re-scatter the incident EM waves with unnatural properties (e.g., negative refraction, perfect absorption, and anomalous reflection). Based on that, LISA is able to enhance transmissions of desired signals and simultaneously to suppress undesired signals. In the following, we review some emerging applications and design strategies of LISA-assisted wireless networks.

\begin{figure}[t]
\centering
\includegraphics[width=5cm]{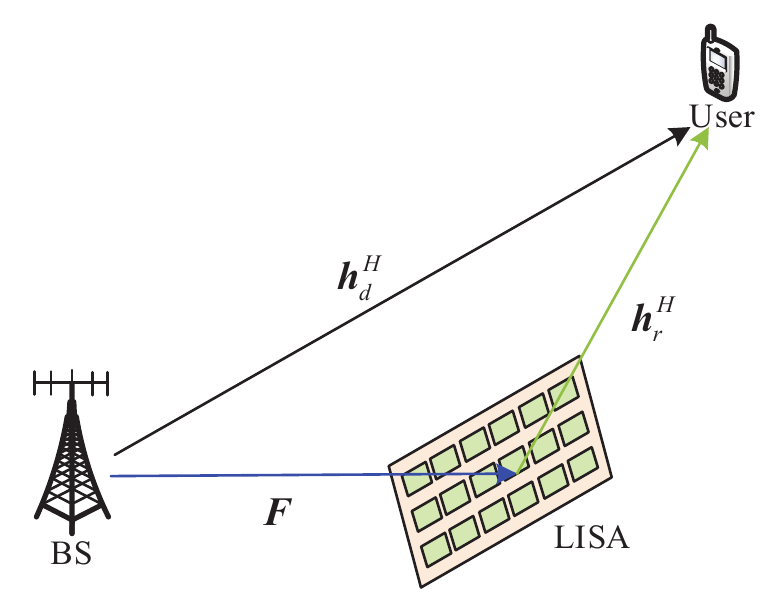}
\caption{System model of downlink transmission in single-user case assisted by LISA.} \label{figureapplication1}
%\vspace{-1.7em}
\end{figure}

\subsection{Downlink Transmission in Single-User Case}
 %point out the practical constraints on the reflecting coefficients.
In Fig. \ref{figureapplication1}, LISA is deployed to enhance the MISO downlink signal transmission system, which consists of one multi-antenna BS, one LISA, and one mobile user. In the following, we first present the signal and the channel model, and then  discuss the channel estimation issue for this novel system.
\subsubsection{Signal and Channel Model}
Denote  the channel responses from the BS to the user, from the LISA to the user, and from the BS to the LISA by ${\bm h}_d$, ${\bm h}_r$, and ${\bm F}$, respectively. The composite BS-LISA-user channel is  usually referred to as the dyadic backscatter channel \cite{griffin2009complete, Ruizhang2018Sufface}. Each element on the LISA combines all the received multi-path signals, and then forwards them to the receiver side resembling a point source by reflection. Thus, the received signal at the user is given by
\begin{eqnarray}
&y= \left[ {{\bm{h}}_{r}^H{\bm \Theta} {\bm{F}} + {\bm{h}}_{d}^H} \right]{{{\bm{w}}}}x  + u,\label{equ:receivedsignal}
\end{eqnarray}
where $\bm w$, $x$, and $u$ are the $N$-dimension downlink beamforming vector, one-dimension transmit signal of the BS, and one-dimension received noise at the user; ${\bm \Theta}={\text {  diag}}([\theta_1,\theta_2,\cdots,\theta_{M}])$ is the $M\times M$-dimension reflection coefficient channel matrix of the LISA, and ${\text { diag}}(\cdot)$ denotes a diagonal matrix whose diagonal elements are given by the corresponding vector.

For this novel system, the key  problem is jointly optimizing the transmit beamforming vector at the BS and the  reflection coefficients at the LISA  to achieve various objects.
Specifically, a received signal power maximization problem was formulated in \cite{Ruizhang2018Sufface}, where  a {\emph{semidefinite relaxation}} (SDR) based iterative optimization approach was developed.
In order to overcome the performance loss of the SDR method caused by the rank relaxation, manifold optimization based method was further adopted in \cite{yu2019miso} to resolve the joint beamforming problem in a more efficient way.
Then, the above model was extended into an {\emph{orthogonal frequency division multiplexing}} (OFDM) system in \cite{yang2019irs}, where the downlink achievable rate was maximized by jointly optimizing power allocation in each subcarrier at the BS and the reflective array reflection coefficients (phases/amplitudes) at the LISA.
Besides, since the {\emph{line-of-sight}} (LoS) signal may be blocked by the obstacles in \emph{millimeter-wave} (mmWave) communication, the LISA with $60$ GHz reflective arrays was implemented to establish robust mmWave connections in \cite{tan2018enabling} , where a three-party beam-searching protocol and the optimal array deployment strategy were developed to minimize the outage probability.

\subsubsection{Channel Estimation Method}\label{sec:channel}
The channel acquisition in the LISA-assisted system is challenging due to the cascaded structure of the dyadic backscatter channel  and the large number of the reflective-radio elements. In particular, based on the signal model in \eqref{equ:receivedsignal}, the received signal power can be rewritten as
\begin{eqnarray}
\left| {({{\bm{h}}_{r}^H{\bm \Theta} {\bm{F}} + {\bm{h}}_{d}^H}){\bm w}} \right|^2=\left| {\bm \theta}^H{\bm H}{\bm w}  \right|^2,
\end{eqnarray}
where ${\bm H}=\left[ {\begin{array}{*{20}{c}}{\text { diag}}({\bm{h}}_{r}^H){\bm F}\\{\bm{h}}_{d}^H\end{array}} \right]$ and ${\bm \theta}=\left[\theta_1,\theta_2,\cdots,\theta_{M},1\right]$. Hence, to obtain the optimal $\bm \theta$ and $\bm w$, we could estimate the composite channel ${\text {diag}}({\bm{h}}_{r}^H){\bm F}$ instead of estimating ${\bm{h}}_{r}$ and ${\bm F}$ independently. Specifically, all channels can be estimated by the following two-stage method:
\begin{itemize}
\item \emph{First stage:} Turn off the LISA, and  pilots are sent from the user to the BS to estimate the direct channel ${{\bm h}_d}$.
\item \emph{Second stage:} Turn on each reflective-radio element on the LISA successively in $N$ time slots, while keeping the other reflective-radio elements closed. Then, in each slot, pilots are sent from the user to the BS to estimate the channel ${{\bm h}_d^H}+{{\bm h}_r^*}(m){\bm F}(m)$, where ${{\bm h}_r^*}(m)$ and ${\bm F}(m)$ are $m$-th element of ${{\bm h}_r^H}$ and $m$-th row vector of ${\bm F}$. Finally, we exploit the estimated ${{\bm h}_d^H}$ to recover ${{\bm h}_r^*}(m){\bm F}(m)$.
\end{itemize}

The above channel estimation method is straightforward, but the overhead of pilots becomes prohibitive if the number of reflective-radio elements is large. To reduce the training overhead, a novel LISA architecture with both passive elements and few active elements was proposed in \cite{taha2019enabling}, in which compressive sensing and deep learning were applied to obtain the CSIs with low complexity and low signal processing Latency.
%Moreover, the ergodic capacity of the studied system and the effects of the phase shifts on the ergodic capacity were investigated  in \cite{han2018large} by exploiting statistical CSI in various propagation scenarios.

\begin{figure}[t]
\centering
\includegraphics[width=5.5cm]{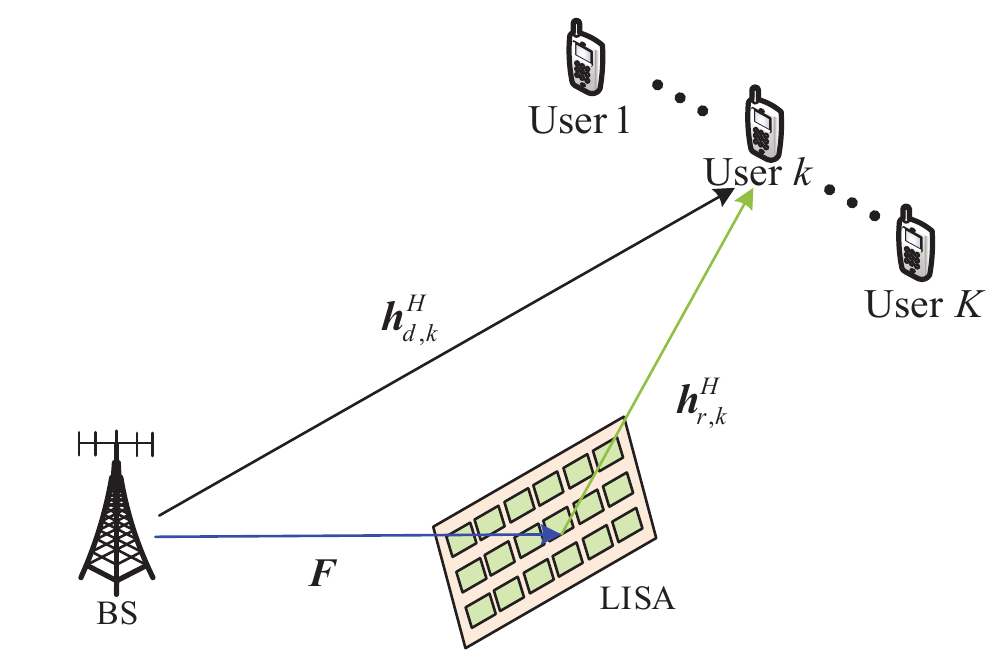}
\caption{System model of downlink transmission in multi-user case assisted by LISA.} \label{figureapplication12}
%\vspace{-1.7em}
\end{figure}

\subsection{Downlink Transmission in Multi-user Case}
The single-user MISO system can be further extended to the multi-user cases, as shown in Fig. \ref{figureapplication12}, which consists of  a multi-antenna BS, one LISA, and $K$ mobile users equipped with single antenna.
Denote the channel responses from the BS to the $k$-th user and from the LISA to the $k$-th user by ${\bm h}_{d,k}$ and ${\bm h}_{r,k}$, respectively. The received signal at the $k$-th user is given by
\begin{eqnarray}
&y_k= \left[ {{\bm{h}}_{r,k}^H{\bm \Theta} {\bm{F}} + {\bm{h}}_{d,k}^H} \right]{\bm s}  + u_k,\label{equ:receivedsignal2}
\end{eqnarray}
where ${\bm s}$ and $u_k$ are the transmitted signal and the received noise at the $k$-th user, respectively. Specifically, we have ${\bm s}=\sum\nolimits_{k = 1}^K {{{\bm{w}}_k}{x_k}} $ for the downlink unicast scenario, i.e., the BS transmits independent signals to each users, and  ${\bm w}_k$ and $x_k$ are the beamforming vector and information signal for the $k$-th user. Similarly,  we have ${\bm s}={\bm{w}}x$ for the downlink multicast scenario, i.e., the BS transmits common signals to all users.

For the downlink unicast scenario, the sum rate and energy-/spectral-efficiency optimization problems were studied  in \cite{huang2018achievable} and \cite{huang2018large} subject to the individual QoS requirement of each user, in which joint alternating maximization algorithm was proposed based on the majorization-minimization method to solve the formulated non-convex problems.
Besides, the weighted sum rate optimization problem was studied in \cite{guo2019weighted}, where fractional programming technique was applied to decouple the non-convex problem, and then an alternating optimization algorithm was developed to solve the problem efficiently.
Furthermore, results from {\emph{random matrix theory}} (RMT) were exploited in \cite{nadeem2019largearxiv} to develop the deterministic approximated solution for the beamforming design, based on which an efficient algorithm was developed by exploiting projected gradient ascent. However, the downlink multicast scenario is still lack of investigation in existing literature.

\subsection{LISA-Assisted Wireless Power Transfer System}
LISA may also be applied to the wireless power transfer system, in which the LISA provides an assisting link for power transfer from the BS to the energy receiver. A novel channel estimation protocol without requiring any prior knowledge of CSI  was proposed in \cite{mishra2019channel}, and near-optimal closed-form expressions for energy beamforming at the BS and the passive beamforming at the LISA were developed to maximize received powers at the intended users.

\begin{figure}[t]
\centering
\includegraphics[width=5.5cm]{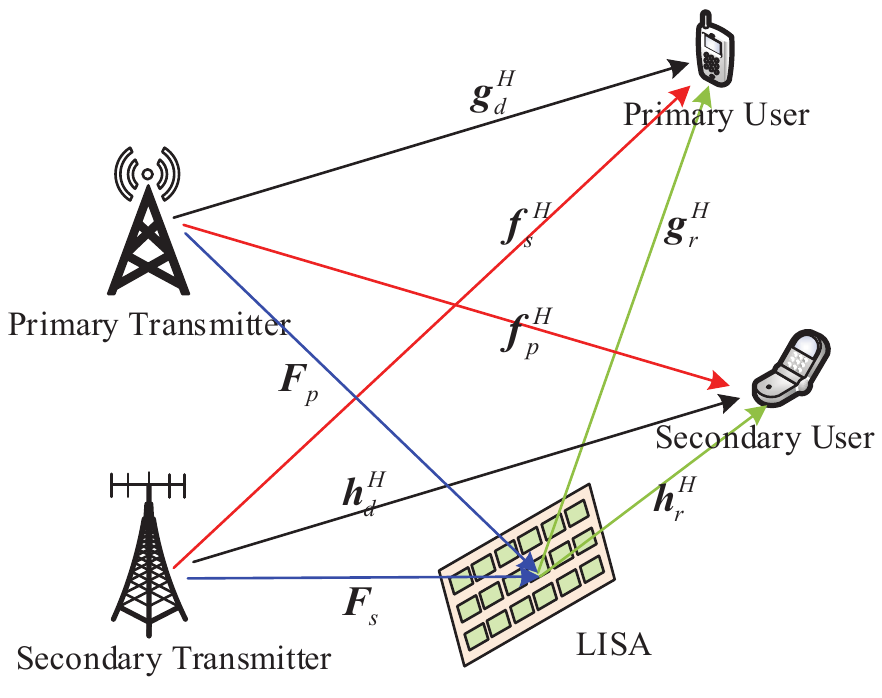}
\caption{System model of cognitive radio network assisted by LISA.} \label{CRN}
%\vspace{-1.7em}
\end{figure}

\subsection{LISA-Assisted Cognitive Radio Network}
In Fig. \ref{CRN}, LISA is deployed in {\emph{cognitive radio network}} (CRN) to enhance transmissions for both primary system and secondary system by suppressing the inter-system interference.
Denote the channel responses from {\emph{primary transmitter}} (PT) to {\emph{primary user}} (PU), from PT to {\emph{secondary user}} (SU), and from PT to LISA by ${\bm{g}}_d$, ${\bm{f}}_p$, and ${\bm{F}}_p$, respectively.
Similarly,  denote the channel from {\emph{secondary transmitter}} (ST) to PU, from ST to SU, and from ST to LISA by ${\bm{h}}_d$, ${\bm{f}}_s$, and ${\bm{F}}_s$, respectively.
Also, denote the channel responses from the LISA to the PU and SU by ${\bm{g}}_r$ and ${\bm{h}}_r$, respectively. Then, the received signals at the PU (i.e., ${y_p}$) and SU (i.e., ${y_s}$) can be further written as
\begin{eqnarray}
{y_p} = {\bm{g}}_d^H{{\bm{s}}_p} + {\bm{f}}_s^H{{\bm{s}}_s} + {\bm{g}}_r^H{\bm \Theta} \left( {{{\bm{F}}_p}{{\bm{s}}_p} + {{\bm{F}}_s}{{\bm{s}}_s}} \right) + {u_p},\\
{y_s} = {\bm{h}}_d^H{{\bm{s}}_s} + {\bm{f}}_p^H{{\bm{s}}_p} + {\bm{h}}_r^H{\bm \Theta} \left( {{{\bm{F}}_p}{{\bm{s}}_p} + {{\bm{F}}_s}{{\bm{s}}_s}} \right) + {u_s},
\end{eqnarray}
respectively, where ${\bm{s}}_p$ and ${\bm{s}}_s$ are the signals transmitted from PT and ST, respectively, ${u_p}$ and ${u_s}$ are the noises at PU and SU, respectively.

In traditional CRN without LISA, one efficient solution to the protection of the primary system as well as achieving high-efficient secondary transmission is adopting beamforming techniques at ST to suppress the interference from ST to PU.
However, the beamforming gain decreases drastically, when PU and SU are allocated in the same directions from ST, such that ${\bm{f}}_s$ and ${\bm{h}}_d$ are highly correlated.
As a result, in these scenarios, it is really hard to guarantee the protection of the primary link while achieving high secondary transmission efficiency simultaneously.
As shown in Fig. \ref{CRN}, by exploiting the LISA, additional virtual links for both primary transmission and secondary transmission can be provided by carefully designing the passive beamforming.
Therefore, the system performances of the primary system and the secondary system will be both improved, especially for the correlated ${\bm{f}}_s$ and ${\bm{h}}_d$ cases. An experimental testbed was developed in \cite{tan2016increasing} to verify the effectiveness of the LISA in CRN.

\subsection{LISA-Assisted Physical Layer Security}
\begin{figure}[t]
\centering
\includegraphics[width=6cm]{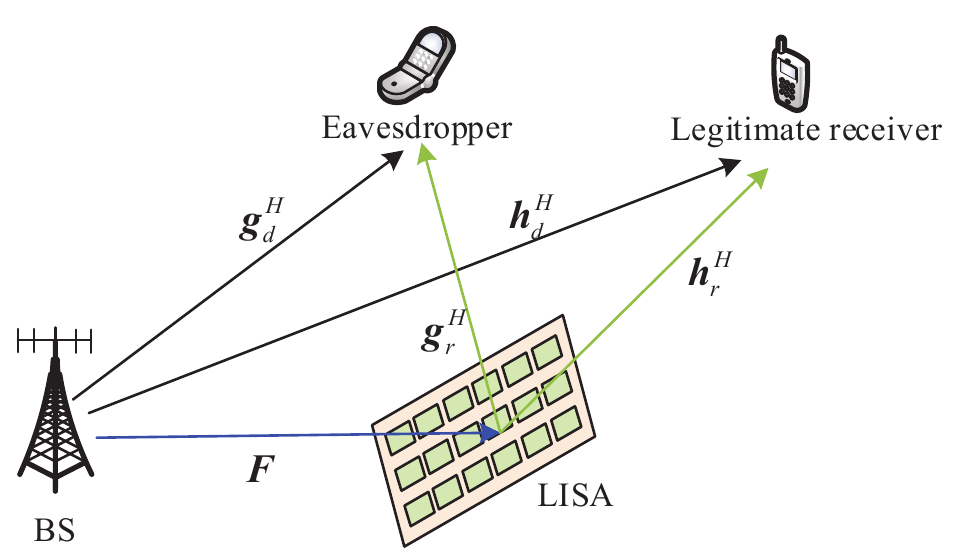}
\caption{System model of secret communication assisted by LISA.} \label{PHY}
%\vspace{-1.7em}
\end{figure}

In Fig. \ref{PHY}, LISA is deployed to assist secret communication, in which the BS transmits confidential data stream to the legitimate receiver against multiple eavesdroppers via the virtual safe secret link through the LISA.
Denote the channel responses from the BS to the LISA, from the BS to the legitimate receiver, from the BS to the eavesdropper, from the LISA to the legitimate receiver, and from the LISA to the eavesdropper by $\bm F$, ${\bm{h}}_d$, ${\bm{g}}_d$, ${\bm{h}}_r$, and ${\bm{g}}_r$, respectively. Then, the received signals at the legitimate receiver (i.e., $y_L$) and eavesdropper (i.e., $y_E$) are given by
\begin{eqnarray}
{y_L} = \left( {{\bm{h}}_d^H + {\bm{h}}_r^H{\bm{\Theta F}}} \right){\bm{w}}s + {u_L},\\
{y_E} = \left( {{\bm{g}}_d^H + {\bm{g}}_r^H{\bm{\Theta F}}} \right){\bm{w}}s + {u_E},
\end{eqnarray}
respectively, where ${\bm{w}}$ is the beamforming vector of the confidential signal $s$, and ${u_E}$ and ${u_L}$ are the received noises at the legitimate receiver and the eavesdropper with powers $\delta_E^2$ and $\delta_L^2$, respectively. From \cite{wyner1975wire}, the secrecy rate  can be written as
\begin{eqnarray}
C &= &\left[ {\log \left( {1 + \frac{1}{{\delta _L^2}}{{\left\| {\left( {{\bm{h}}_d^H + {\bm{h}}_r^H{\bm{\Theta F}}} \right){\bm{w}}} \right\|}^2}} \right)} \right.\nonumber\\
&&\quad{\left. { - \log \left( {1 + \frac{1}{{\delta _E^2}}{{\left\| {\left( {{\bm{g}}_d^H + {\bm{g}}_r^H{\bm{\Theta F}}} \right){\bm{w}}} \right\|}^2}} \right)} \right]^ + },
\label{eqPHY1}\end{eqnarray}
where  $[a]^{+}\!=\!{\rm max}(0,a)$. Similarly to the CR cases, it is intractable to guarantee the secret communication by only applying beamforming techniques,  if the legitimate receiver and the eavesdropper are allocated in the same directions to the BS for the highly correlated channel responses ${\bm{h}}_d$ and ${\bm{g}}_d$.
Hopefully, the confidential data stream may bypass the eavesdropper through the LISA to reach the legitimate receiver, so that the secrecy rate is improved.
Based on this idea, a minimum-secrecy-rate maximization problem among multiple eavesdroppers was formulated in \cite{chen2019IRSPHY} under various practical constraints on the reflective-radio elements, which capture the scenarios of both continuous and discrete reflective coefficients of the reflective-radio elements. In addition, an efficient algorithm based on alternating optimization and a path-following algorithm was developed to solve the formulated non-convex problem efficiently.

\section{Limitations, Challenges, and Open Research Problems}
\label{sec:chall}

LISA has attracted much attention since it makes the wireless environment more smart to improve resource utilization. However, there are some critical limitations, challenges, and open research problems in LISA, which are discussed as follows.

\subsection{Channel Characterization}

When LISA are integrated into the systems, it introduces different channel properties, e.g. channel correlation and reflecting effect brought by metasurface, and so on. To assess the actual performance, it is necessary to have a more accurate channel model, which reflect the true physical behavior of the radio channel.
In addition, channel estimation with LISA is more difficult than that without LISA since there exist multiplicative channel and the reflective-radio elements as shown in \eqref{equ:receivedsignal}. It is relatively easy to estimate the entirety of the multiplicative channel, which is discussed in Section \ref{sec:channel}, while it is difficult to separate it into two channels if the LISA does not have the ability to receive signal. However, the knowledge of the two separated channels is very important for LISA to assist transmission. Thus the estimation of the two separated channels is a challenging open problem.
Furthermore, it is necessary to consider how imperfect channel information affects the system design and how to adjust the LISA to assist transmission under imperfect channel information.

\subsection{LISA Deployment}

One significant design aspect while integrating LISA into wireless networks is the deployment problem. For LISA-aided communications, appropriate deployment of LISA can significantly improve the system performance. In particular, when there are LoS paths between the transmitter and LISA, and between the LISA and receiver. Thus, where to deploy LISA and how many LISA units need to be deployed are important problems to be solved.

\subsection{Network Optimization and Resource Allocation}

When the environment can be software-defined, the network optimization and resource allocation schemes are changed since the LISA can assist the information transmission for the users with poor channel gain which need not to choose the other channels or access other points.
Thus, it is very important to design new network optimization and resource allocation schemes. In addition, when a residential area is deployed many LISAs, it is a very important yet open problem that how the LISAs cooperate and interact with each other to achieve a global optimum.

\subsection{Artificial Intelligence}

Artificial intelligence technology is an indispensable part to make the environment smart. One possible solution is that the BS controls all LISAs in one cell in assisting multi-user transmission using AI techniques. On the other hand, a LISA can be viewed as an independent agent to make decisions or optimize the environment by observing behaviors and incoming signals. That is to say, multiple LISAs make decisions independently with distributed learning by interacting with each other to achieve a global optimum.

\subsection{Security and Privacy}

Although the environment can be software-defined with LISA, security and privacy are also critical problems in future communication with LISA since the environment can be the target of adversarial attacks. For example, attackers could disrupt the LISAs through the wireless channel to make them out of control and lead to misbehavior. In addition, a selfish user served by LISA could feedback fake information to the agent to manipulate the LISA in order to enhance its communication performance. That is, the use of LISA leads to new challenges in terms of security and privacy.
Therefore, it is an urgent problem to design a suitable and effective policy to ensure the security and privacy for communication systems with LISA.

\section{Conclusion}\label{conclusion}
In this paper, we have provided a comprehensive overview of the reflective radio basics and the promising LISA technology.
Implementing LISA in wireless communication networks opens an emerging area for intentionally programming the wireless environment to improve communication quality.
In particular, we highlight various potential applications based on LISA, such as reflective massive MIMO, wireless power transfer, cognitive radio, physical layer security, and symbiotic radio network.
As a new emerging technique, there are still many open issues still requiring significant research efforts to make the benefit of LISA a reality. We hope this work become a useful and effective guidance for future work in this area.

\bibliographystyle{IEEEtran}%By using IEEEtrans, the number can be displayed.
%\bibliography{ref_AmBC}
\bibliography{IEEEabrv,mybib_draft2}

\end{document}